\documentclass[twocolumn,appendixfloats]{emulateapj}
\usepackage{float}
\usepackage{epsfig}
\usepackage{graphicx}
\usepackage{graphics}
\usepackage[latin1]{inputenc}
\usepackage{latexsym}
\usepackage{url}
\usepackage{multirow}
\slugcomment{Accepted for publication in The Astronomical Journal}
\shorttitle{NIR Cepheid P-L and P-W relations}
\shortauthors{Bhardwaj et al.}

\begin{document}

\title{Large Magellanic Cloud Near-Infrared Synoptic Survey. II.\\
The Wesenheit relations and their application to the Distance scale}

\author{Anupam Bhardwaj\altaffilmark{1,*}, Shashi M. Kanbur\altaffilmark{2}, Lucas M. Macri\altaffilmark{3},
Harinder P. Singh\altaffilmark{1}, Chow-Choong Ngeow\altaffilmark{4},\\ R. Wagner-Kaiser\altaffilmark{5} \& Ata Sarajedini\altaffilmark{5}}

\altaffiltext{1}{Department of Physics \& Astrophysics, University of Delhi, Delhi 110007, India. }
\altaffiltext{2}{Department of Physics, The State University of New York at Oswego, Oswego, NY 13126, USA.}
\altaffiltext{3}{Mitchell Institute for Fundamental Physics \& Astronomy, Department of Physics \& Astronomy, 
Texas A\&M University, College Station, TX 77843, USA.}
\altaffiltext{4}{Graduate Institute of Astronomy, National Central University, Jhongli 32001, Taiwan.}
\altaffiltext{5}{University of Florida, Department of Astronomy, 211 Bryant Space Science Center, Gainesville, FL, 32611.}
\altaffiltext{*}{Corresponding author; {\tt anupam.bhardwajj@gmail.com}}

\begin{abstract}
We present new near-infrared Cepheid Period-Wesenheit relations in the LMC using time-series observations 
from the Large Magellanic Cloud Near-Infrared Synoptic Survey. 
We also derive optical$+$near-infrared P-W relations using $V$ and $I$~magnitudes from OGLE-III.
We employ our new $JHK_s$ data to determine an independent distance to the LMC of $\mu_{\rm LMC} = 18.47\pm0.07
{\textit{(statistical)}}$~mag,
using an absolute calibration of the Galactic relations based on several distance determination methods and accounting 
for the intrinsic scatter of each technique. We also derive new near-infrared Period-Luminosity and Wesenheit relations 
for Cepheids in M31 using observations from the PHAT survey. We use the absolute calibrations of the Galactic and LMC $W_{J,H}$ 
relations to determine the distance modulus of M31, $\mu_{\rm M31} = 24.46\pm0.20$~mag. We apply a simultaneous 
fit to Cepheids in several Local Group galaxies covering a range of metallicities ($7.7<12+\log[O/H]<8.6$~dex) to determine 
a global slope of -$3.244\pm0.016$~mag/dex for the $W_{J,K_s}$ relation and obtain robust distance estimates. Our distances
are in good agreement with recent TRGB based distance estimates and we do not find any evidence for a metallicity 
dependence in the near-infrared P-W relations.

\end{abstract}


\section{Introduction}

Studies of Cepheid variables are of considerable interest in determining distances 
to star-forming galaxies out to $\sim 50$~Mpc because these pulsating stars
obey the well known Period-Luminosity relation or Leavitt Law \citep{levitt12} and hence can be used as
standard candles. In the era of precision cosmology, Cepheids play an important role in the
cosmic distance scale and are vital in establishing an increasingly more accurate and
precise value of the Hubble constant \citep{riess09, riess11}. In the recent past, 
many studies have used classical Cepheids as standard candles for cosmic 
distance scale work through the Period-Luminosity (P-L) and Period-Luminosity-Color (PLC) relations 
\citep{bono99, tammann03, smk03, sandage04, persson04, benedict07, sandage09}. Most of these studies 
involve the calibration of P-L relations for the Galaxy and LMC at optical wavelengths.
Some authors assume that the Galactic and LMC P-L relations have similar slopes 
\citep{fouque07, monson12}. However, the universality of Cepheid P-L relation
is a subject of intense debate as the metallicity and extinction effects might change the slope as well as the
intercept of the P-L relation \citep{gieren06, storm11} and therefore lead to biases in distance determinations. 

Near-infrared (NIR) Cepheid P-L relations acquire a greater significance because these are less susceptible to reddening and metallicity 
differences between target and calibrating galaxies \citep{storm11, monson12}. Another possible reason for 
discrepancy in Cepheid-based distance estimates 
is the significant non-linearities at various periods during the different
phases of pulsation at optical wavelengths \citep{ngeowsmk06, anupam14}. These non-linearities are also observed for
mean light P-L relations at optical bands but are expected to be less significant at NIR wavelengths 
\citep{bono99, madore09}.

The calibration of Galactic Cepheid P-L relations at optical and NIR bands has been carried out 
using parallaxes for small samples of variables \citep{tammann03, ngeowsmk04, benedict07, fouque07, turner10, storm11}.
For example, \citet{benedict07} used highly accurate trigonometric parallaxes from the Hubble Space Telescope 
(HST) for 10 Cepheids. The major problem in obtaining solid calibrations 
within our Galaxy is that accurate distance determinations are only possible for nearby objects  
\citep[$D\lesssim 500$~pc with HST/FGS, recently extended to $D\lesssim 4$~kpc with a ``spatial scanning technique'' by][]{riess14}. 
The most important fundamental distance measurements come from trigonometric parallaxes.
The Hipparcos/Tycho catalogues of parallaxes for Classical Cepheids gave a strong impetus to this field \citep{Perryman97, van07}.
Cepheid distances have also been measured to high precision by the Infrared Surface Brightness technique and
Baade-Wesselink methods, where Cepheid pulsation is directly measured with a long-baseline interferometer 
\citep{gieren98, storm11, groen13}.

Recently, a detailed study on Period-Wesenheit (P-W) relations in the NIR bands was carried out
to determine distances to the Magellanic Clouds by \citet{inno13}. The reddening-free Wesenheit function \citep{madore82} in the
optical bands was also used to derive distances to individual Galactic Cepheids \citep{ngeow12}. The author 
calibrated the P-L relations at both optical and infrared wavelengths and used these to determine a distance
modulus to the LMC. At NIR wavelengths, \citet{persson04} derived the P-L relations for Cepheids
in the LMC having full phased light curve data and determined the distance modulus to the LMC using Galactic calibrations
from the literature.

Determining a robust distance to the LMC is an important step in the cosmic distance scale. Recently, \citet{piet13}
used a sample of 8 eclipsing binaries to obtain a 2.2\% accurate distance to the LMC of
$D = 49.97\pm1.11~kpc$ (equivalent to $\mu_{\rm LMC} = 18.493\pm0.048$~mag).
One of the motivations for our work is to provide an independent determination of the LMC distance modulus by applying
a Galactic calibration to data from the Large Magellanic Cloud NIR Synoptic Survey (LMCNISS) \citep[][hereafter Paper I]{macri15}.
We also extend the distance determination to M31 using recent observations for Cepheids from the PHAT 
survey \citep{rachel15}. Our work also provides a test for the metallicity dependence of Cepheid based distance estimates, 
considering the fact that Local Group galaxies have a large metallicity range ($7.7<12+\log[O/H]<8.6$~dex). 
Further, this work will be especially 
important in light of the impending launch of the {\it James Webb Space Telescope} (JWST) in a few years, 
when space-based observations of Cepheids will be exclusively available in the infrared bands. 
A robust absolute calibration of the NIR P-L relations for Cepheids in the Milky Way 
and LMC will play an important role in the cosmic distance scale.

This paper, the second in a series, is structured as follows. In Section~\ref{sec:lmcplr}, 
we present the absolute P-W relations for Cepheids in the LMC using data from Paper I. We 
determine the robust distance to the LMC using absolute calibration of 
the Galactic Cepheid P-L and P-W relations (\S3). We also derive the P-L and P-W relations 
for M31 using the observations from the PHAT survey \citep{dal12, william14} in Section~\ref{sec:m31plr}.
Finally, we use Galactic and LMC calibrations to determine metal-independent robust distances to 
Local Group galaxies (\S5). Further discussion of the results and important conclusions of our study
are presented in Section~\ref{sec:discuss}.

\section{NIR Period-Wesenheit Relations for the LMC Cepheids}
\label{sec:lmcplr}

\subsection{Photometric mean magnitudes}

We make use of NIR mean magnitudes for 789 fundamental-mode and 475 first-overtone Cepheids 
in the LMC from Paper I. These magnitudes are based on observations from a synoptic survey (average of 16 epochs) 
of the central region of the LMC using the CPAPIR camera at the Cerro Tololo Interamerican Observatory (CTIO) 
1.5-m telescope between 2006 and 2007. Most of these Cepheid variables were previously studied in the optical $V$ and $I$ 
bands by the third phase of Optical Gravitational Lensing Experiment (OGLE-III) survey \citep{sosz2008, ulac13}.
The $V$ and $I$ band mean magnitudes are also compiled in Paper I.  
The calibration into the 2MASS photometric system, extinction corrections and the adopted reddening 
law are discussed in detail in Paper I.

\subsection{Absolute Calibration of NIR P-W Relations}

We derive new NIR and optical+NIR P-W relations for fundamental and first-overtone mode Cepheids using LMCNISS and OGLE data. 
We note that Paper I presents only the P-L relations;
therefore, it is important to derive P-W relations for their application to the distance scale.
Moreover, we also emphasize that this large homogeneous data set in the $JHK_s$ bands 
for Cepheids in the LMC are based on time-series observations as opposed to single-phase as in earlier studies.
We modify the definition of the Wesenheit function
relative to \citet{inno13} as:

\begin{eqnarray}
\label{eq:pw_all}
W^{\lambda_{3}}_{\lambda_{2},\lambda_{1}} & = & m_{\lambda_{3}} - R^{\lambda_{2},\lambda_{1}}_{\lambda_3} (m_{\lambda_{2}}-m_{\lambda_{1}}), \\
R^{\lambda_{2},\lambda_{1}}_{\lambda_3} & = & \left[\frac{A_{\lambda_{3}}}{E(m_{\lambda_{2}}-m_{\lambda_{1}})}\right] \nonumber \\
\end{eqnarray}

\noindent where $m_{\lambda_i}$ represents the mean magnitude at wavelength $\lambda_i$ and $\lambda_{1}>\lambda_{2}$.
For simplicity, the superscript $\lambda_3$ is dropped from $W$ when $\lambda_1 = \lambda_3$. We adopt the reddening law given in \citet{card89} and assume a value of $R^{B,V}_V = 3.23$ to
obtain selective absorption ratios A$_{I}$/A$_{V}$ = 0.610, A$_{J}$/A$_{V}$ = 0.292, A$_{H}$/A$_{V}$ = 0.181,
\& A$_{K_{s}}$/A$_{V}$ = 0.119 \citep{fouque07, inno13}. The resulting Wesenheit relations studied in this work are listed in Table~\ref{table:pw_all}.

\begin{table}
\begin{center}
\caption{Wesenheit relations}
\label{table:pw_all}
\scalebox{1.0}{
\begin{tabular}{llcc}
\hline
\hline
Label  		&	 $m_{\lambda_3}$		&	 $R^{\lambda_2,\lambda_1}_{\lambda_3}$ & $m_{\lambda_2} - m_{\lambda_1}$   \\
\hline
W$_{J,H}$    & H   & 1.63 & J-H   \\
W$_{J,K_s}$  & K$_s$ & 0.69 & J-K$_s$ \\
W$_{H,K_s}$  & K$_s$ & 1.92 & H-K$_s$ \\
W$_{V,J}$    & J   & 0.41 & V-J   \\
W$_{V,H}$    & H   & 0.22 & V-H   \\
W$_{V,K_s}$  & K$_s$ & 0.13 & V-K$_s$ \\
W$_{I,J}$    & J   & 0.92 & I-J   \\
W$_{I,H}$    & H   & 0.42 & I-H   \\
W$_{I,K_s}$  & K$_s$ & 0.24 & I-K$_s$ \\
W$^H_{V,I}$  & H   & 0.41 & V-I   \\
\hline
\end{tabular}}
\end{center}
\end{table}

\begin{table*}
\begin{center}
\caption{Wesenheit magnitudes for Cepheids in the LMC}
\label{table:lmc_wmag}
\scalebox{1.0}{
\begin{tabular}{cccccccccccccccccc}
\hline
\hline
Star ID & Type & $\log P$ & $W_{J,H}$ & $W_{J,K_s}$ & $W_{H,K_s}$ & $W_{V,J}$ & $W_{V,H}$ & $W_{V,K_s}$ & $W_{I,J}$ & $W_{I,H}$ & $W_{I,K_s}$ & $W^H_{V,I}$\\
&&&$\sigma_{W_{J,H}}$ & $\sigma_{W_{J,K_s}}$ & $\sigma_{W_{H,K_s}}$ & $\sigma_{W_{V,J}}$ & $\sigma_{W_{V,H}}$ & $\sigma_{W_{V,K_s}}$ & $\sigma_{W_{I,J}}$ & $\sigma_{W_{I,H}}$ & $\sigma_{W_{I,K_s}}$& $\sigma_{W^H_{V,I}}$ \\
\hline
0477&FO&      0.292  &    13.922&    14.238&    14.471&    14.732&    14.820&    14.407&    14.853&    14.351&    14.403&    14.397     \\ &   &  &      0.132&     0.085  &     0.173&     0.058&     0.053&     0.065&     0.082&     0.062&     0.067&     0.061\\
0478&FU&      0.442  &    14.124&    14.354&    14.523&    14.497&    14.649&    14.408&    14.533&    14.314&    14.404&    14.371     \\ &   &   &      0.167&     0.102  &     0.219&     0.064&     0.059&     0.079&     0.089&     0.082&     0.081&     0.079\\
0482&FU&      0.873  &    12.520&    12.820&    13.042&    13.296&    13.494&    12.988&    13.386&    12.921&    12.974&    13.006     \\ &   &   &      0.142&     0.066  &     0.088&     0.088&     0.080&     0.031&     0.117&     0.040&     0.033&     0.042\\
0487&FU&      0.493  &    13.930&    14.093&    14.215&    14.528&    14.663&    14.244&    14.590&    14.235&    14.228&    14.305     \\ &   &   &      0.212&     0.139  &     0.212&     0.132&     0.122&     0.092&     0.171&     0.049&     0.094&     0.050\\
0488&FU&      0.562  &    13.805&    14.088&    14.296&    14.271&    14.484&    14.158&    14.349&    14.057&    14.160&    14.102     \\ &   &   &      0.104&     0.085  &     0.169&     0.044&     0.039&     0.068&     0.067&     0.053&     0.070&     0.054\\

\hline
\end{tabular}}
{\tablecomments{All 789 fundamental and 475 first-overtone mode Cepheids were used to derive NIR Wesenheit relations, while
$3\sigma$ clipping was applied for optical+NIR relations. The uncertainties were calculated by
propagating the errors in mean magnitudes.}}
\end{center}
\end{table*}

\begin{figure*}
\begin{center}
\includegraphics[width=0.8\textwidth,keepaspectratio]{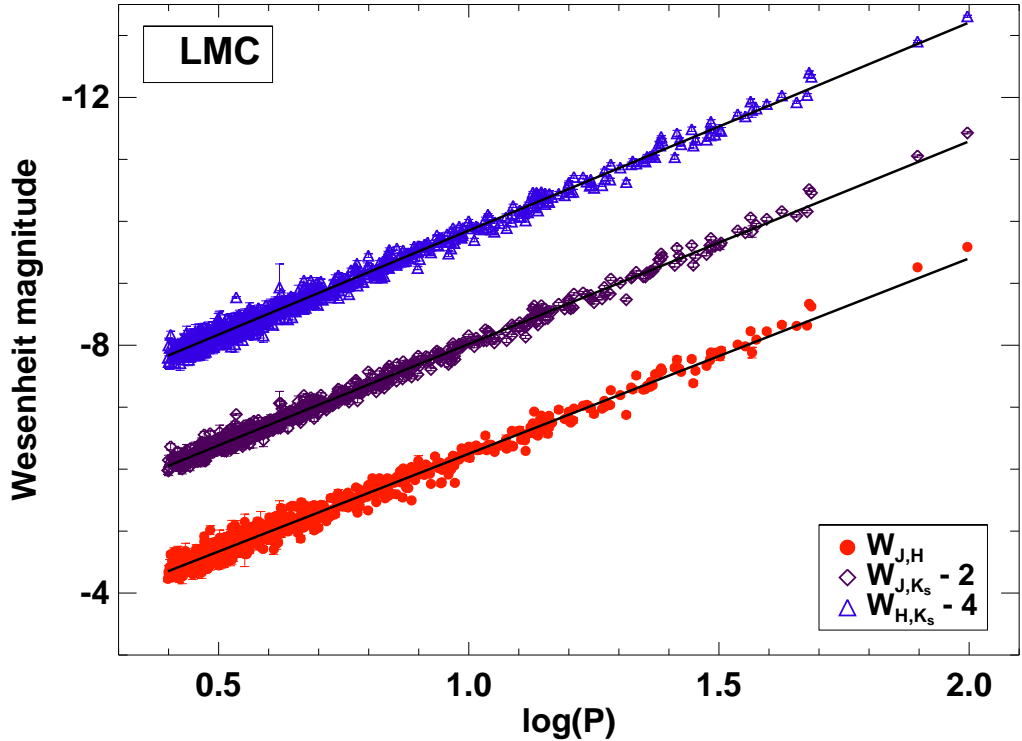}
\caption{Calibrated NIR P-W relations for fundamental-mode Cepheids in the LMC. The solid line represents 
the best-fit linear regression to the data points in each band.}
\label{fig:pw_lmc.eps}
\end{center}
\end{figure*}

\begin{figure*}
\begin{center}
\includegraphics[width=0.8\textwidth,keepaspectratio]{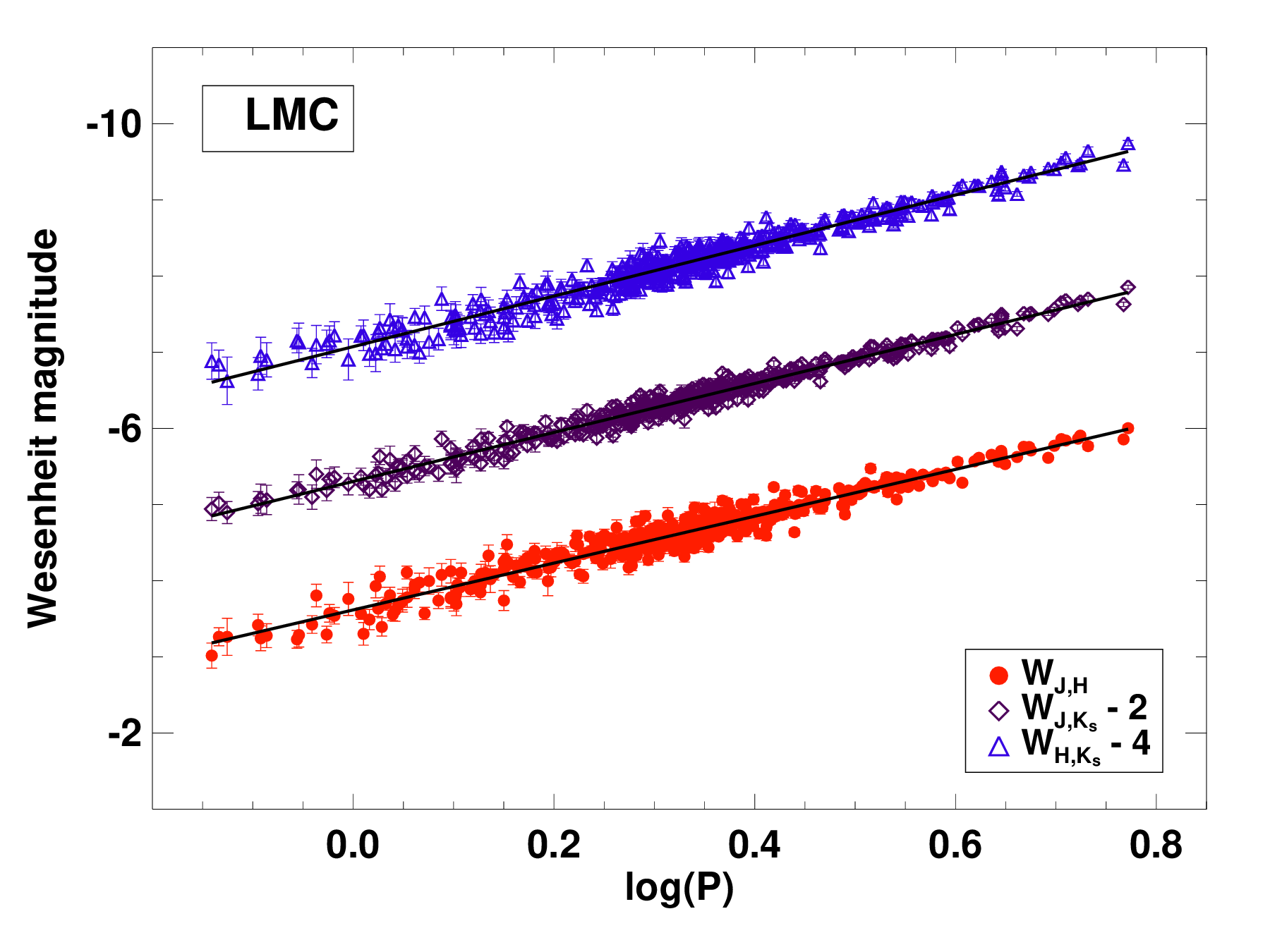}
\caption{Calibrated NIR P-W relations for first-overtone mode Cepheids in the LMC. The solid line represents 
the best-fit linear regression to the data points.}
\label{fig:fo_pw_lmc.eps}
\end{center}
\end{figure*}

\begin{figure*}
\begin{center}
\includegraphics[width=0.99\textwidth,keepaspectratio]{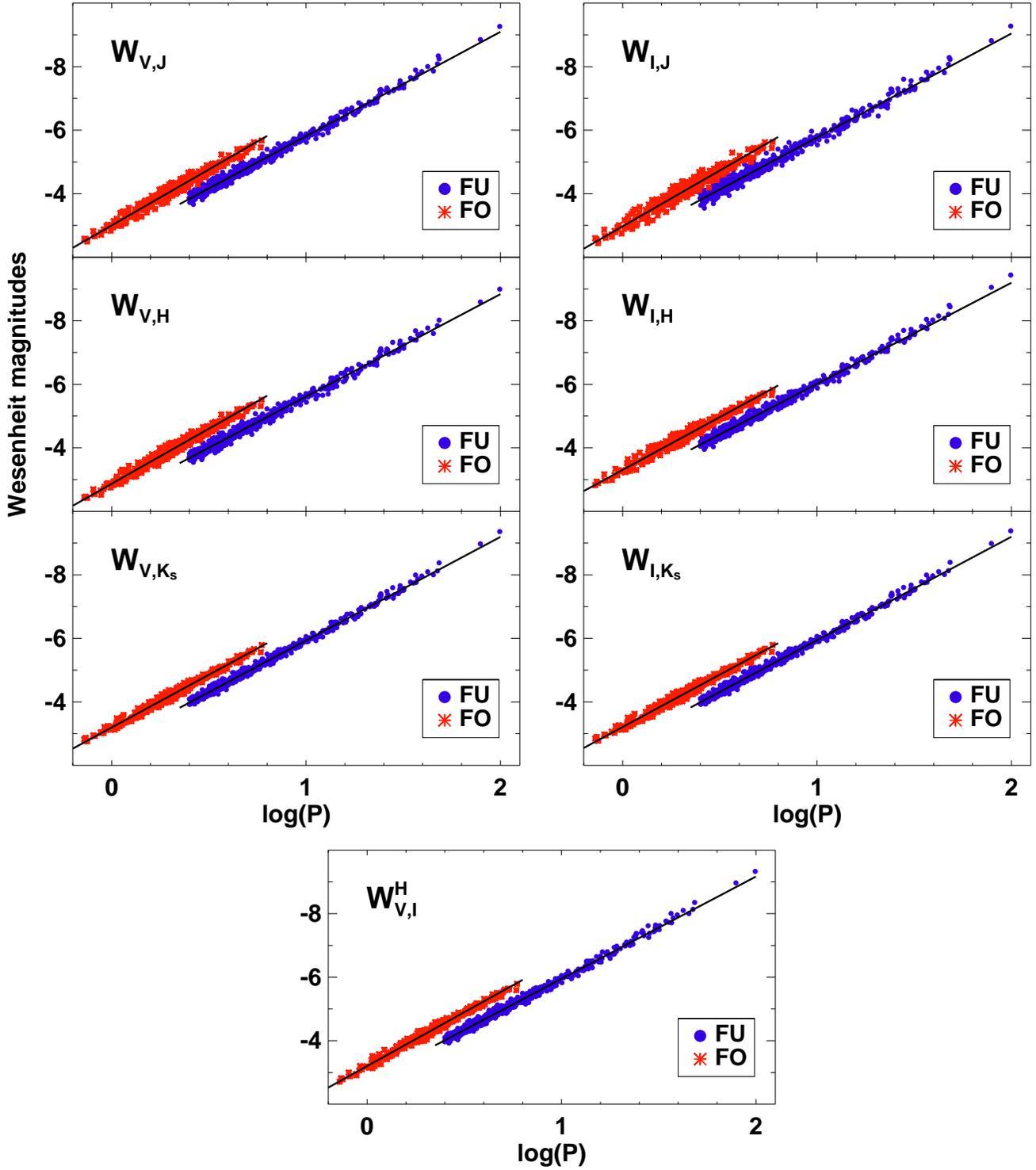}
\caption{Calibrated optical+NIR P-W relations for fundamental and first-overtone mode Cepheids in the LMC. 
The solid line represents the best-fit linear regression to the data points.}
\label{fig:opt_nir.eps}
\end{center}
\end{figure*}

The Wesenheit magnitudes are given in Table~\ref{table:lmc_wmag}, together with their propagated uncertainties. 
For the NIR relations, we use the final sample of Cepheids 
from Paper I, since sigma-clipping was already applied in that work.
Following Paper I, we calibrate these Wesenheit magnitudes 
using the highly accurate LMC distance from \citet{piet13}. 
The calibrated Wesenheit magnitudes for fundamental and first-overtone 
mode Cepheids are plotted separately against $\log(P)$ to fit a P-W relation in the form of 
$W_{\lambda_{2},\lambda_{1}} = a[\log(P) - 1] + b$. The results 
for the fundamental and first-overtone mode Cepheids in the LMC are 
shown in Figs.~\ref{fig:pw_lmc.eps} and \ref{fig:fo_pw_lmc.eps},
respectively. In the case of optical+NIR Wesenheit relations, we apply $3\sigma$ 
clipping to the magnitudes before fitting a P-W relation. The optical+NIR 
P-W relations for fundamental and first-overtone Cepheids are shown in
Fig.~\ref{fig:opt_nir.eps}, with the derived parameters given
in Table~\ref{table:lmc_slope}. We also include a calibration of 
the $W^H_{V,I}$ relation, which is the primary method used by the 
SH0ES project \citep{riess09,riess11} to determine Cepheid distances 
to SNe Ia hosts and ultimately estimate the Hubble constant.

We also provide the P-L relations from Paper I in Table~\ref{table:lmc_slope} for relative comparison with the P-W 
relations and the Galactic P-L relations in the next sections. Previously, the largest set of full phased light curve data 
used in the calibration of the NIR P-L and P-W relations consisted of a sample of only 92 stars from \citet{persson04}. 
However, this data set includes a larger number of stars with periods between $10-100$ days which were used
in the Paper I and this analysis for the determination of the NIR P-L and P-W 
relations, respectively. We also list in Table~\ref{table:lmc_slope} the LMC $K_s$ P-L relation
and the $W_{V,K_s}$ P-W relation 
derived by \citet{ripepi12} based on data from the VISTA survey of the Magellanic Clouds System (VMC).

The reddening-free Wesenheit relations are expected to have a smaller dispersion than the corresponding P-L relations.
We note from Table~\ref{table:lmc_slope} that the P-L relations for fundamental-mode Cepheids in $J$ and $H$ show a dispersion 
(0.125 and 0.103 mag) slightly greater than K$_{s}$ (0.09 mag). For Wesenheit relations, this
dispersion reduces significantly in $W_{J,H}$ and $W_{J,K_{s}}$ as compared to $J$ and $H$. 
In the case of $W_{H,K_{s}}$, the dispersion increases relatively as compared to the K$_{s}$, presumably
due to an insignificant contribution from the colour ($H-K_{s}$) term.  For first-overtone Cepheids,
$W_{J,K_s}$ relation has the smallest dispersion as compared to dispersion in $J$, $H$ and $K_s$ (0.134, 0.100 and 0.086) 
P-L relations. Similarly, the $W_{J,H}$ and $W_{H,K_{s}}$ Wesenheit also show smaller
dispersions similar to fundamental mode P-L relations. These P-W relations play a vital role 
in determining reddening independent accurate distances \citep{inno13}.

\begin{table*}
\begin{center}
\caption{LMC Cepheid NIR P-L and P-W relations}
\label{table:lmc_slope}
\scalebox{1.05}{
\begin{tabular}{p{0.7cm}ccccccccc}
\hline
\hline
 & Slope & Intercept & $\sigma$ &  N & Src & \multicolumn{2}{c}{Slope} & \multicolumn{2}{c}{Intercept}\\
  &	&		&	&	&	&	$|T|$	&	$p(t)$	&	$|T|$	&$p(t)$	\\
\hline
\hline
   \multicolumn{10}{c}{fundamental-mode} \\
\hline
	  $J$& 	   -3.140$\pm$0.004 	&     -5.308$\pm$0.051     &      0.125&     789&M15&---&---&---&---\\       
	  $H$&     -3.169$\pm$0.004 	&     -5.674$\pm$0.053     &      0.103&     789&M15&---&---&---&---\\
        $K_s$&     -3.228$\pm$0.004     &     -5.737$\pm$0.048     &      0.090&     789&M15&---&---&---&---\\
             &     -3.295$\pm$0.018     &     -5.718$\pm$0.051     &      0.102&     256&R12&      3.96&       0.00&      0.28&      0.78\\
    $W_{J,H}$&     -3.154$\pm$0.014     &     -6.248$\pm$0.049     &      0.108&     789&TW&---&---&---&---\\
             &     -3.373$\pm$0.008     &     -6.236$\pm$0.048     &      0.080&    1701&I13&     14.89&       0.00&      0.18&      0.86\\
$W_{J,K_{s}}$&     -3.272$\pm$0.011     &     -6.020$\pm$0.049     &      0.078&     789&TW&---&---&---&---\\
             &     -3.365$\pm$0.008     &     -5.982$\pm$0.048     &      0.080&    1708&I13&      6.78&       0.00&      0.55&      0.58\\
$W_{H,K_{s}}$&     -3.358$\pm$0.013     &     -5.851$\pm$0.049     &      0.101&     789&TW&---&---&---&---\\
             &     -3.360$\pm$0.010     &     -5.795$\pm$0.048     &      0.100&    1709&I13&      0.12&       0.90&      0.81&      0.42\\
    $W_{V,J}$&     -3.303$\pm$0.012     &     -5.815$\pm$0.049     &      0.092&     708&TW&---&---&---&---\\
             &     -3.272$\pm$0.009     &     -5.787$\pm$0.048     &      0.080&    1732&I13&      2.15&       0.03&      0.42&      0.68\\
    $W_{V,H}$&     -3.239$\pm$0.013     &     -5.618$\pm$0.049     &      0.095&     711&TW&---&---&---&---\\
             &     -3.315$\pm$0.008     &     -5.992$\pm$0.048     &      0.070&    1730&I13&      5.46&       0.00&      5.56&      0.00\\
$W_{V,K_{s}}$&     -3.284$\pm$0.010     &     -5.943$\pm$0.049     &      0.073&     709&TW&---&---&---&---\\
             &     -3.326$\pm$0.008     &     -5.918$\pm$0.048     &      0.070&    1737&I13&      3.32&       0.00&      0.37&      0.71\\
             &     -3.325$\pm$0.014     &     -5.948$\pm$0.050     &      0.078&     256&R12&      2.44&       0.01&      0.07&      0.94\\
    $W_{I,J}$&     -3.290$\pm$0.016     &     -5.774$\pm$0.049     &      0.115&     715&TW&---&---&---&---\\
             &     -3.243$\pm$0.011     &     -5.734$\pm$0.049     &      0.100&    1735&I13&      2.53&       0.01&      0.59&      0.56\\
    $W_{I,H}$&     -3.226$\pm$0.012     &     -6.027$\pm$0.049     &      0.088&     711&TW&---&---&---&---\\
             &     -3.317$\pm$0.008     &     -6.009$\pm$0.048     &      0.080&    1734&I13&      6.53&       0.00&      0.27&      0.79\\
$W_{I,K_{s}}$&     -3.281$\pm$0.010     &     -5.952$\pm$0.049     &      0.076&     710&TW&---&---&---&---\\
             &     -3.325$\pm$0.008     &     -5.916$\pm$0.048     &      0.070&    1737&I13&      3.52&       0.00&      0.53&      0.59\\
  W$^H_{V,I}$&     -3.249$\pm$0.010     &     -5.958$\pm$0.048     &      0.076&     710&TW&---&---&---&---\\

\hline
   \multicolumn{10}{c}{first-overtone mode} \\
\hline
          $J$&     -3.297$\pm$0.020     &     -6.986$\pm$0.050     &      0.134&     475&M15&---&---&---&---\\
          $H$&     -3.215$\pm$0.020     &     -6.252$\pm$0.050     &      0.100&     475&M15&---&---&---&---\\
        $K_s$&     -3.245$\pm$0.023     &     -6.304$\pm$0.050     &      0.086&     475&M15&---&---&---&---\\
             &     -3.471$\pm$0.035     &     -6.384$\pm$0.049     &      0.099&     256&R12&      5.62&       0.00&      1.15&      0.25\\
    $W_{J,H}$&     -3.075$\pm$0.035     &     -6.692$\pm$0.050     &      0.118&     475&TW&---&---&---&---\\
             &     -3.507$\pm$0.015     &     -6.793$\pm$0.048     &      0.090&    1064&I13&     12.76&       0.00&      1.49&      0.14\\
$W_{J,K_{s}}$&     -3.216$\pm$0.024     &     -6.520$\pm$0.049     &      0.082&     475&TW&---&---&---&---\\
             &     -3.471$\pm$0.013     &     -6.594$\pm$0.048     &      0.080&    1057&I13&      9.45&       0.00&      1.08&      0.28\\
$W_{H,K_{s}}$&     -3.319$\pm$0.035     &     -6.393$\pm$0.050     &      0.119&     475&TW&---&---&---&---\\
             &     -3.425$\pm$0.017     &     -6.435$\pm$0.049     &      0.100&    1063&I13&      2.94&       0.00&      0.62&      0.54\\
    $W_{V,J}$&     -3.436$\pm$0.029     &     -6.457$\pm$0.049     &      0.095&     422&TW&---&---&---&---\\
             &     -3.434$\pm$0.014     &     -6.452$\pm$0.048     &      0.100&    1086&I13&      0.06&       0.95&      0.07&      0.94\\
    $W_{V,H}$&     -3.390$\pm$0.028     &     -6.275$\pm$0.049     &      0.093&     421&TW&---&---&---&---\\
             &     -3.485$\pm$0.011     &     -6.621$\pm$0.048     &      0.080&    1071&I13&      3.42&       0.00&      5.16&      0.00\\
$W_{V,K_{s}}$&     -3.293$\pm$0.021     &     -6.493$\pm$0.049     &      0.071&     421&TW&---&---&---&---\\
             &     -3.456$\pm$0.013     &     -6.539$\pm$0.048     &      0.070&    1061&I13&      6.64&       0.00&      0.67&      0.50\\
             &     -3.530$\pm$0.025     &     -6.623$\pm$0.049     &      0.070&     256&R12&      7.24&       0.00&      1.89&      0.06\\
    $W_{I,J}$&     -3.433$\pm$0.036     &     -6.425$\pm$0.050     &      0.118&     420&TW&---&---&---&---\\
             &     -3.423$\pm$0.020     &     -6.417$\pm$0.048     &      0.130&    1100&I13&      0.23&       0.82&      0.11&      0.91\\
    $W_{I,H}$&     -3.254$\pm$0.026     &     -6.573$\pm$0.049     &      0.086&     422&TW&---&---&---&---\\
             &     -3.489$\pm$0.012     &     -6.631$\pm$0.048     &      0.080&    1072&I13&      8.52&       0.00&      0.86&      0.39\\
$W_{I,K_{s}}$&     -3.279$\pm$0.021     &     -6.493$\pm$0.049     &      0.074&     420&TW&---&---&---&---\\
             &     -3.448$\pm$0.013     &     -6.539$\pm$0.048     &      0.080&    1059&I13&      6.60&       0.00&      0.66&      0.51\\
  W$^H_{V,I}$&     -3.313$\pm$0.021     &     -6.533$\pm$0.050     &      0.070&     421&TW&---&---&---&---\\

\hline
\end{tabular}}
{\tablecomments{Source: TW - this work; M15 - \citet{macri15}; R12 - \citet{ripepi12};
I13 - \citet{inno13}. The intercepts of the P-L and P-W relations from R12 and I13 were transformed to the 2MASS system
and recast as $M_\lambda = a_\lambda[\log(P) - 1] + b_\lambda$ for ease of comparison.}}
\end{center}
\end{table*}

\begin{table*}
\begin{center}
\caption{Fourier-fitted mean magnitudes}
\label{table:mean_mag}
\scalebox{1.0}{
\begin{tabular}{cccccccccc}
\hline
\hline
Star & Source & $P$  & \multicolumn{3}{c}{Magnitudes ($m_{0}$)} & \multicolumn{3}{c}{$\sigma(m_{0})$} & $E(B-V)$ \\
ID	&   &(days)		&$J$	&$H$	&$K_{s}$	&$J$	&$H$	&$K_{s}$& \\		
\hline
AK CEP & MP & 7.233 & 8.408 & 7.888 & 7.741 & 0.022 & 0.024 & 0.025 & 0.635 \\
AN AUR & MP & 10.291 & 7.934 & 7.436 & 7.275 & 0.022 & 0.024 & 0.026 & 0.600 \\
AQ PUP & LS & 30.104 & 6.001 & 5.491 & 5.308 & 0.023 & 0.021 & 0.022 & 0.531 \\
AW PER & MP & 6.464 & 5.229 & 4.822 & 4.697 & 0.022 & 0.024 & 0.025 & 0.487 \\
BB SGR & LS & 6.637 & 5.053 & 4.641 & 4.512 & 0.045 & 0.021 & 0.022 & 0.276 \\
\hline
\end{tabular}}
\tablecomments{Source: MP - \citet{monson11}; BTG - \citet{barnes97}; LS - \citet{laney92}; W - \citet{welch84}. The colour excess $E(B-V)$ values are taken from 
\citet{tammann03}. The error estimate includes the uncertainties from Fourier fit and the photometry. This table is 
available entirely in a machine-readable form in the online journal. Only first five lines are shown here for guidance 
regarding its form and content.}
\end{center}
\end{table*}

\subsection{Comparison with published LMC P-L and P-W relations}
We also compare our P-W relations in the LMC with \citet{ripepi12} and \citet{inno13}. 
We use a standard $t$-test to check the consistency of the slopes and intercepts of our P-L and P-W relations
with published work. Under the null hypothesis that the two slopes are equivalent, the T-values are calculated by incorporating
errors on the slopes and the standard deviation. The theoretical $t_{\alpha/2,\nu}$ values are evaluated from the 
$t$-distribution, where we adopt the significance level of $\alpha=0.05$ and $\nu=N_1 + N_2 - 4$ with $N_1$ and $N_2$ being
the number of Cepheids in the two samples. The probability ($p(t)$) of the observed t-statistic ($|T|$) under the null 
hypothesis is listed in Table~\ref{table:lmc_slope}. The theoretical t-value, at a fixed $\alpha$, varies marginally
($\sim$1.96 - 1.98) for a wide range of $\nu$ (100 - 3000) used in our calculations and therefore, is not tabulated.
The null hypothesis is rejected if $|T|>t$ or $p(t) < 0.05$ 
i.e. the slopes or zero-points are not equal.

We find that the slope of our $K_s$-band P-L relation for fundamental and first-overtone mode 
Cepheids is not consistent with the slope of the P-L relation from the VMC survey \citep{ripepi12}. 
However, the intercepts are statistically consistent between these two studies.
Our slopes for the fundamental-mode NIR P-W relations are statistically different from those of
\citet{inno13} in $W_{J,H}$ and $W_{J,K_{s}}$, while being consistent in $W_{H,K_{s}}$.
Similarly, the slopes for all optical+NIR P-W relations are not consistent with the results of \citet{inno13}.
In the case of the first-overtone mode Cepheids, the slope of the NIR P-W relations
from this study are significantly different from the results of \citet{inno13}, while for 
the optical+NIR P-W relations, only the $W_{V,J}$ and $W_{I,J}$ P-W relations have 
similar slopes. However, the intercepts of most P-W relations for both fundamental and first-overtone mode Cepheids are 
in good agreement, given their uncertainties. The t-test also suggests
that the zero-points of our relations are statistically consistent with previously published results, 
except in case of $W_{V,H}$. The possible reason for inconsistency in slopes may be 
due to significantly different sample sizes and different photometric calibrations.  
Moreover, the mean magnitudes in \citet{inno13} are obtained from a template fit to single-epoch magnitudes for 
fundamental-mode Cepheids, while random-phase magnitudes are used for first-overtone Cepheids. 
Therefore, we emphasize that all our results are based on mean magnitudes from well-sampled light curves.


\section{An Independent Distance to the LMC using Galactic P-L and P-W Relations}
\label{sec:Galacticplr}

A precise determination of the distance to the LMC is essential to estimate a value of
Hubble constant with a total uncertainty below 2\% \citep{riess09, riess11}. We aim to determine an
independent and robust distance to the LMC using Cepheids as standard candles, following 
the work of \citet{piet13} based on long-period, late-type eclipsing binaries.
The Wesenheit and $JHK_s$~magnitudes from this work and Paper I, respectively, can also be used to obtain an 
independent estimate of distance to the LMC. An additional feature of this approach is the use of 
mean magnitudes based on full phased NIR light curves in the target galaxy (the LMC) as opposed 
to corrected single-epoch observations. However, this requires an absolute calibration of the P-L and P-W relations 
in the Galaxy. Previous Galactic P-L relations vary significantly in slope and zeropoint,
leading to differences of more than $\sim3\%$ in the inferred LMC distance. 
A detailed comparison of distance estimates to the LMC using published Galactic P-L relations is provided in 
appendix~\ref{sec:comp_mw_lmc}. Therefore, we re-analysed the available data in the literature to provide 
a new robust absolute calibration of the Galactic
relations, as explained in the following subsections. 

\subsection{Absolute Calibration of NIR Galactic Relations}

We make use of light curve data for 113 Galactic Cepheids in the $JHK_{\rm s}$ bands from the literature 
\citep{welch84, laney92, barnes97, monson11} for which independent distances are available. The light 
curve data for these Cepheids along with their Fourier analysis is discussed in detail in \citet{bhardwaj14}. 
The mean magnitudes obtained using the optimum-order Fourier fit \citep{barts82, bhardwaj14} along with their errors are listed in 
Table~\ref{table:mean_mag}. We compare the Fourier-fitted mean magnitudes with values from the literature and the difference 
between two sets do not exceed $\sim0.02$~mag. Since the NIR light curve data compiled from various sources are in 
different photometric systems, we converted these mean magnitudes to the 
2MASS photometric system using the standard colour 
transformations\footnote{\tiny \url{$http://www.ipac.caltech.edu/2mass/releases/allsky/doc/sec6\_4b.html$}}.
This transformation led to an average change in color of $\sim0.02$~mag.
In order to obtain reddening-corrected mean magnitudes in all the three bands, colour excess $E(B-V)$ for 
Galactic Cepheids are adopted from \citet{tammann03}. We adopt the \citet{card89} reddening law as discussed previously and
use the absorption ratios to determine R$_{J}$ = 0.94, R$_{H}$ = 0.58 and R$_{K}$ = 0.38.
We adopt an uncertainty in the color excess 
equal to the difference between two independent determinations of $E(B-V)$ for all of these Cepheids, 
$\Delta E(B-V)\sim 0.03$~mag \citep{fernie95}, and propagate this uncertainty into the errors in mean magnitudes 
using equations given in \citet{tammann03}. 

\begin{table*}
\begin{center}
\caption{Galactic Cepheid Distance Moduli}
\label{table:adopted_dis}
\scalebox{1.0}{
\begin{tabular}{cccccccccccc}
\hline
\hline
Star ID & IRSB & $\sigma$(IRSB) & MS & $\sigma$(MS) & BW & $\sigma$(BW) & HST-$\pi$ & $\sigma$(HST-$\pi$) & W.M. ($\mu$) & $\sigma(\mu)$  \\
\hline
AK CEP& ---& ---& ---& ---& 13.03& 0.20& ---& ---& 13.03& 0.20\\
AN AUR& ---& ---& ---& ---& 13.62& 0.22& ---& ---& 13.62& 0.22\\
AQ PUP& 12.53& 0.04& 11.78& 0.10& 12.38& 0.06& ---& ---& 12.40& 0.63\\
AW PER& ---& ---& ---& ---& 9.94& 0.18& ---& ---& 9.94& 0.18\\
BB SGR& 9.69& 0.03& 9.08& 0.08& 9.55& 0.07& ---& ---& 9.58& 0.51\\
\hline
\end{tabular}}
{\tablecomments{The distance determination methods :  Hubble Space Telescope parallaxes 
(HST-$\pi$) \citep{benedict07, monson12, riess14}, Infrared Surface Brightness (IRSB) method 
\citep{fouque07, storm11}, Baade-Wesselink (BW) distances \citep{groen13}, main-sequence (MS) fitting to candidate 
cluster \citep{turner10}. We provide the distance moduli compiled from various methods 
for relative comparison. The adopted distance modulus is the weighted mean (W.M.) of all available distance moduli 
for each star. The procedure adopted to estimate uncertainties listed in the last column is discussed in the text.
This table is available entirely in a machine-readable form in the online journal. Only first five lines are shown here for guidance 
regarding its form and content.}}
\end{center}
\end{table*}

\begin{figure}
\begin{center}
\includegraphics[width=0.95\columnwidth,keepaspectratio]{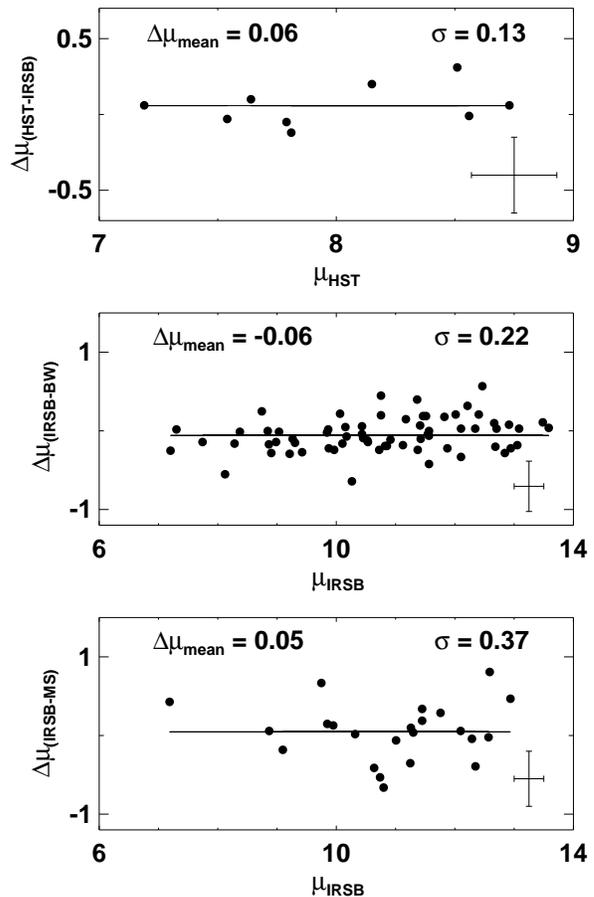}
\end{center}
\caption{Comparisons of distance moduli obtained using different techniques. The solid lines represent 
the mean value and representative error bars show the corresponding median uncertainties from Table~\ref{table:adopted_dis}.}
\label{fig:offset_dis.eps}
\end{figure}

\begin{table*}
\begin{center}
\caption{Calibrated Magnitudes for Fundamental-Mode Galactic Cepheids}
\label{table:abs_mag}
\scalebox{1.0}{
\begin{tabular}{cccccccccccccccc}
\hline
\hline
Star & $\log(P)$ & \multicolumn{3}{c}{Absolute magnitudes} & \multicolumn{3}{c}{$\sigma$(Absolute mag.)} &F$_{1}$  & \multicolumn{3}{c}{Wesenheit magnitudes} & \multicolumn{3}{c}{$\sigma$(Wesenheit mag.)} &F$_{2}$\\
ID & - & M$_{J}$ &  M$_{H}$ &  M$_{K}$ & M$_{J}$ &  M$_{H}$ &  M$_{K}$ & - & $W_{J,H}$ & $W_{J,K}$ & $W_{H,K}$ & $W_{J,H}$ & $W_{J,K}$ & $W_{H,K}$ & -\\
\hline
AK CEP & 0.859 & -5.223 & -5.561 & -5.573 & 0.201 & 0.201 & 0.201 & Y & -6.112 & -5.814 & -5.596 & 0.208 & 0.202 & 0.210 &Y \\
AN AUR & 1.012 & -6.254 & -6.529 & -6.615 & 0.221 & 0.222 & 0.221 & N & -6.978 & -6.864 & -6.780 & 0.229 & 0.222 & 0.231 &N \\
AQ PUP & 1.479 & -6.955 & -7.201 & -7.312 & 0.630 & 0.630 & 0.630 & Y & -7.604 & -7.558 & -7.524 & 0.631 & 0.630 & 0.631 &Y \\
AW PER & 0.810 & -5.181 & -5.452 & -5.471 & 0.181 & 0.181 & 0.181 & Y & -5.893 & -5.671 & -5.508 & 0.189 & 0.183 & 0.191 &Y \\
BB SGR & 0.822 & -4.837 & -5.128 & -5.190 & 0.510 & 0.510 & 0.510 & Y & -5.603 & -5.434 & -5.310 & 0.511 & 0.510 & 0.512 &Y \\

\hline
\end{tabular}}
{\tablecomments{The uncertainties in absolute magnitudes include the errors in mean magnitudes and distance moduli from 
Table~\ref{table:mean_mag} and \ref{table:adopted_dis}, errors from transformations to 2MASS system and reddening 
corrections. These errors are propagated to estimate uncertainty for Wesenheit magnitudes. The flags F$_{1}$ and 
F$_{2}$ indicate if the Cepheid is used in final P-L and P-W fits, respectively. This table is 
available entirely in a machine-readable form in the online journal. Only first five lines are shown here for guidance 
regarding its form and content.}}
\end{center}
\end{table*}

We compiled distances from various distance determination methods to calibrate the P-L and P-W relations for Galactic Cepheids: Hubble Space Telescope parallaxes (HST-$\pi$), 
Infrared Surface Brightness (IRSB), Baade-Wesselink (BW) and main-sequence (MS) fitting to candidate 
clusters. Highly accurate HST parallaxes for 11 Galactic Cepheids are available in the literature 
\citep{benedict07, monson12, riess14}. We use the updated values of HST-$\pi$ for BETA DOR and W SGR 
from Table 5 of \citet{monson12}, which differ slightly from those tabulated in \citet{benedict07}.
The Galactic P-L relations based on IRSB, BW, MS distances are discussed in 
\citep{fouque07, turner10, storm11, monson12, groen13}. We note that the principle of 
distance determination using IRSB and BW methods is similar but with 
different treatment of algorithms. \citet{groen13} essentially used the same data as \citet{storm11} 
and hence they are not totally independent of each other. Both these studies found a similar dependence 
of the p-factor on period, but the zero point implied a shorter distance scale. The LMC distance modulus
found by \citet{groen13} was shorter as compared to recent studies. Therefore, we only make use of BW distances 
when the corresponding IRSB distance is not available. The distance moduli from all available
methods for a given Cepheid are listed in Table~\ref{table:adopted_dis}.

\begin{table*}
\begin{center}
\caption{Galactic Cepheid NIR P-L and P-W relations}
\label{table:compare_slope}
\scalebox{1.0}{
\begin{tabular}{p{0.7cm}ccccccccc}
\hline
\hline
  & Slope & Intercept & $\sigma$ &  N & Src & \multicolumn{2}{c}{Slope} & \multicolumn{2}{c}{Intercept}\\
  &	&		&	&	&	&	$|T|$	&	$p(t)$	&	$|T|$	&$p(t)$	\\
\hline
        
   $J$  & -3.127$\pm$0.076 & -5.320$\pm$0.023 & 0.223 &	99 & TW		&---	&---	&---	&---\\
	& -3.194$\pm$0.068 & -5.258$\pm$0.020 & 0.155 & 59 & F07	&      0.60 &      0.55&      1.87 &      0.06 \\
	& -3.180$\pm$0.090 & -5.220$\pm$0.030 & 0.220 & 70 & S11  	&      0.45 &      0.65&      2.64 &      0.01\\
	& -3.058$\pm$0.021 & -5.340$\pm$0.019 & 0.073 & 203 & N12  	&      1.11 &      0.27&      0.52 &      0.61\\
       
   $H$  & -3.164$\pm$0.074 & -5.643$\pm$0.022 & 0.219 & 99 & TW  \\
	& -3.328$\pm$0.060 & -5.543$\pm$0.020 & 0.146 & 56 & F07 	&      1.57 &      0.12&      3.00 &      0.00 \\
	& -3.300$\pm$0.080 & -5.590$\pm$0.030 & 0.220 & 70 &  S11 	&      1.25 &      0.21&      1.43 &      0.16\\
	& -3.181$\pm$0.022 & -5.648$\pm$0.020 & 0.077 & 203 & N12 	&      0.27 &      0.78&      0.13 &      0.90\\
       
$K_{s}$ & -3.278$\pm$0.073 & -5.716$\pm$0.022 & 0.219 & 99 & TW 	&---	&---	&---	&---\\
	& -3.365$\pm$0.062 & -5.647$\pm$0.019 & 0.144 & 58 & F07 	&      0.82 &      0.41&      2.13 &      0.03\\
	& -3.330$\pm$0.090 & -5.660$\pm$0.030 & 0.220 & 70 & S11  	&      0.45 &      0.65&      1.51 &      0.13\\
	& -3.231$\pm$0.021 & -5.732$\pm$0.020 & 0.075 & 203 & N12 	&      0.78 &      0.44&      0.40 &      0.69\\
\hline
$W_{J,H}$  &     -3.223$\pm$0.076      &    -6.168$\pm$0.023      &     0.228&	99 & TW	&---	&---	&---	&---\\
$W_{J,K_s}$&     -3.383$\pm$0.074      &    -5.989$\pm$0.022      &     0.223&	99 & TW	&---	&---	&---	&---\\
	   &     -3.415$\pm$0.074      &    -6.037$\pm$0.071      &     0.230&	70 & S11&0.31	&0.76	&0.66	&0.51\\
$W_{H,K_s}$&     -3.499$\pm$0.075      &    -5.856$\pm$0.023      &     0.225&	99 & TW	&---	&---	&---	&---\\
\hline
\end{tabular}}
{\tablecomments{ The P-L relations are taken from the sources : TW - This work, F07 - \citet{fouque07}, 
S11 - \citet{storm11}, N12 - \citet{ngeow12}. The P-L and P-W relations from some of these studies are 
transformed to the notation of $M = a[\log(P) - 1] + b$ for ease of comparison.}}
\end{center}
\end{table*}

Fig.~\ref{fig:offset_dis.eps} shows comparisons of distance moduli obtained using different techniques. 
We consider HST parallaxes to be highly precise measurements that include realistic estimates of statistical 
and systematic sources of uncertainty (median error of 0.14 mag). In contrast, we note that the values listed 
in Table~\ref{table:adopted_dis} for the uncertainties in BW, IRSB \& MS distance moduli, as reported in the 
original publications, are not consistent with the observed dispersions seen in Fig.~\ref{fig:offset_dis.eps}.
Therefore, we use the latter to estimate a minimum uncertainty for each of these three techniques. Initially, 
we homogenize the sample by correcting each distance from methods other than HST-$\pi$ for average shifts to match HST-$\pi$ distances.
The average shifts between any two methods are ($\Delta$(HST-$\pi$ - IRSB)=0.06, $\Delta$(HST-$\pi$ - BW)=0.10,
$\Delta$(IRSB - BW)=-0.06, $\Delta$(IRSB - MS)=0.05). The BW and IRSB methods are very similar and have the highest 
number of Cepheids in common and also have equal dispersion ($\sigma=0.13$) with HST-$\pi$. We consider an equal 
contribution from each to the variance in the middle panel (Fig.~\ref{fig:offset_dis.eps}) and determine 
their minimum uncertainty as 0.15 mag. We subtract the contribution of IRSB from the observed variance in 
the bottom panel (Fig.~\ref{fig:offset_dis.eps}) to determine a minimum 
error of 0.33 mag for MS distances. We adopt these values as the minimum allowed uncertainty for a given 
technique when calculating the mean error-weighted distances and uncertainties listed in the last column 
of Table~\ref{table:adopted_dis}. For these uncertainties, we adopt a conservative approach and use 
the greater of the standard deviation of the data and the uncertainty on the mean.

\begin{figure*}
\begin{center}
\includegraphics[width=0.75\textwidth,keepaspectratio]{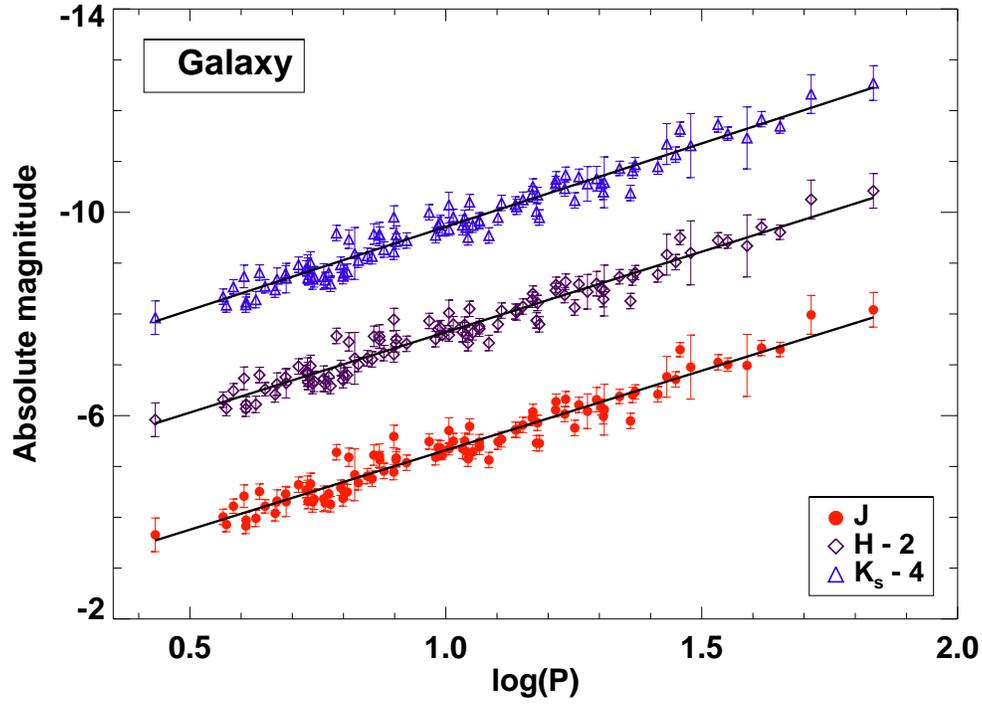}
\caption{Calibrated NIR P-L relations for fundamental-mode Galactic Cepheids. The solid line represents 
the best-fit linear regression to the data points in each band.}
\label{fig:pl_mw.eps}
\end{center}
\end{figure*}

We use extinction-corrected 2MASS mean magnitudes and the adopted mean distance modulus 
given in Table~\ref{table:adopted_dis} to calibrate our Galactic P-L and P-W relations.
The calculated absolute magnitude for each fundamental mode Cepheid is 
presented in Table~\ref{table:abs_mag}. The uncertainty in the absolute magnitude is mostly 
driven by the large uncertainties on distance and also to a lesser extent, on reddening correction errors.
Since our sample included 10 first-overtone stars 
(DT CYG, FF AQL, FN AQL, EV SCT, QZ NOR, SU CAS, SZ TAU, V496 AQL, X LAC, Y OPH) as identified from 
\citet{ngeow12}, we did not consider these stars in calibrating the 
P-L relations. We also restricted our sample to include only those stars that 
have periods greater than 2.5 days. Further, we remove 3$\sigma$ outliers in each NIR band to 
fit a P-L relation, for a final sample of 99 stars.
Absolute magnitudes are plotted against $\log(P)$ and we fit 
a P-L relation in the form, $M_{\lambda} = a_{\lambda}[\log(P) - 1] + b_{\lambda}$,
where $a_{\lambda}$ is the slope and $b_{\lambda}$ is the intercept at $\log(P) = 1$. 
The P-L relations for Galactic Cepheids in each NIR 
band are shown in Fig.~\ref{fig:pl_mw.eps} and the slopes and 
intercepts are given in Table~\ref{table:compare_slope}. 

\begin{figure*}
\begin{center}
\includegraphics[width=0.75\textwidth,keepaspectratio]{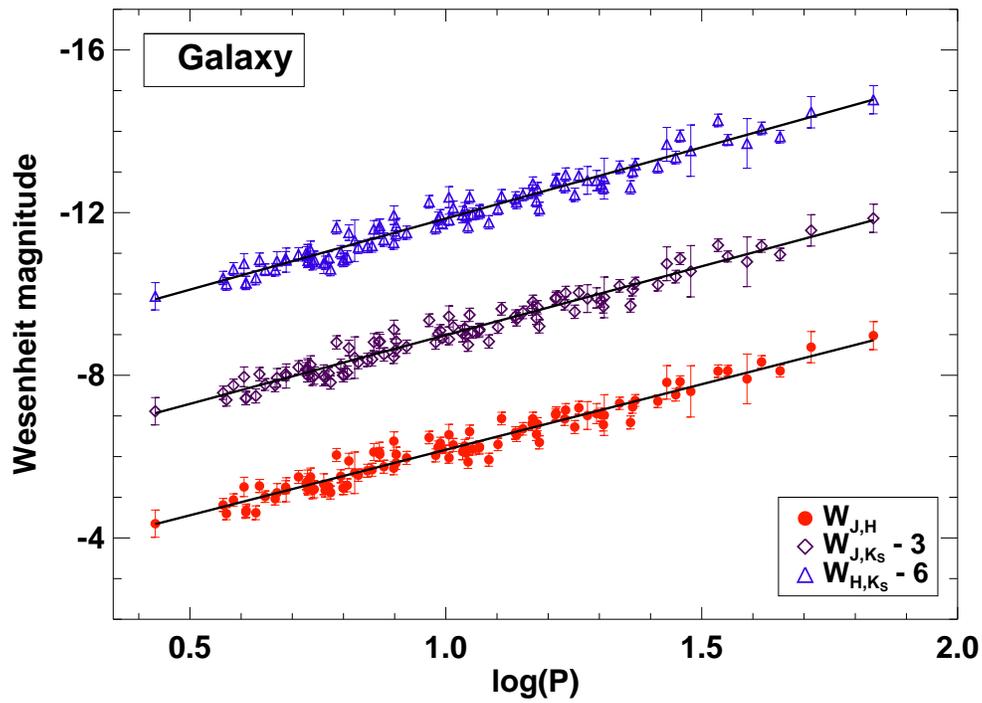}
\caption{Calibrated NIR P-W relations for fundamental-mode Galactic Cepheids. The solid line represents 
the best-fit linear regression to the data points in each band.}
\label{fig:pw_mw.eps}
\end{center}
\end{figure*}

We make use of these calibrated absolute magnitudes to derive P-W relations for the Galaxy. 
These Wesenheit magnitudes are estimated using equation~\ref{eq:pw_all} and given in Table~\ref{table:abs_mag}
together with the absolute magnitudes. We again remove 3$\sigma$ outliers when fitting each P-W relation.
These calibrated P-W relations for the Galaxy are shown in Fig.~\ref{fig:pw_mw.eps} and the results are 
presented in Table~\ref{table:compare_slope}. We compare our Galactic NIR P-L relations with those 
published by \citet{fouque07}, \citet{storm11} and \citet{ngeow12}. The results from these studies are also listed in 
Table~\ref{table:compare_slope} and the detailed comparison is discussed in appendix~\ref{sec:comp_mw_lmc}.

\subsection{Distance to the LMC}
Once the Galactic P-L relation is calibrated we can use it to derive the distance moduli of LMC Cepheids.
Assuming the $JHK_{s}$ P-L relations to have universal slopes and intercepts, we
calculate the absolute magnitude in all bands for individual LMC Cepheids having period $P$.
We have the mean apparent magnitudes from Paper I for all Cepheids in the LMC, and using the calibrated 
absolute magnitudes, we estimate individual distance moduli for all LMC Cepheids. We remove the 
$3\sigma$ outliers in the calculated distance moduli and adopt the average value
to be the distance modulus in each NIR band. In Section~\ref{sec:comp_mw_lmc}, we have provided evidence that
the P-L and P-W relations are universal for the Galaxy and the LMC. Hence, we do not observe any significant trend 
as a function of period in the distance moduli for LMC Cepheids.

The values of mean distance moduli for LMC Cepheids are provided in Table~\ref{table:results_mu}.
These results are in excellent agreement with the result from \citet{piet13}, given their uncertainties. 
This further suggests that the metallicity correction is not needed in $H$ and $K_{s}$ 
and the zero point of the P-L relation in these bands is metallicity independent. 
The reason for the small variations in these distance moduli could be due to the 
slight difference in the slopes and intercepts of the two galaxies and also the
errors in the transformations of $JHK_{s}$ Galactic mean magnitudes to the 2MASS system. 
The LMC distance moduli obtained using the calibration of 
Galactic P-W relations are also presented in Table~\ref{table:results_mu}. 
Again, these distance moduli are in excellent agreement with the \citet{piet13} result. 

The errors in the P-L based distance estimates are only 3\%, while those based on P-W are 4\%. 
We expect that with the larger number of Cepheids having high quality light curve data in the 
LMC OGLE-IV survey, the errors can be reduced further. The Galactic calibrations in our work
are based on distances obtained by four independent methods, which have different sources of systematic 
errors. At present, it is difficult to provide an absolute calibration of Galactic relations with a 
well-determined systematic uncertainty, which can be propagated to Cepheid based distance estimates. 
Therefore, we only provide the total statistical uncertainty and the systematic errors are expected to 
be of the order of, or even larger than the quoted uncertainties. A robust calibration of Galactic
relations will only be possible with accurate parallaxes from {\textit {GAIA}} and then the LMCNISS data can be 
used to obtain a more precise distance to the LMC. However, our results do provide a useful check on
the distance to the LMC, which is consistent and independent to the distance obtained by \citet{piet13}.

\begin{table}
\begin{center}
\caption{LMC distance moduli}
\label{table:results_mu}
\scalebox{1.0}{
\begin{tabular}{cccc}
\hline
\hline
  		&	 $J$		&	 $H$ 		&	$K_{s}$   \\
$\mu_{\rm LMC}$     & 18.52$\pm$0.06 		& 18.47$\pm$0.06 		& 18.47$\pm$0.06\\
\hline
  		&	 $W_{J,H}$		&	 $W_{J,K_{s}}$ 	&	$W_{H,K_{s}}$   \\
$\mu_{\rm LMC}$     & 18.40$\pm$0.08 		& 18.44$\pm$0.09 		& 18.46$\pm$0.09\\
\hline
& \multicolumn{3}{c}{Fixed Galactic P-L slopes}\\
$\mu_{\rm LMC}$     & 18.51$\pm$0.06 		& 18.46$\pm$0.06 		& 18.48$\pm$0.06\\
\hline
\multicolumn{4}{c}{Average value = 18.47$\pm$0.07}\\
\hline
\end{tabular}}
\end{center}
\end{table}

Alternatively, we also calculate the LMC distance moduli using the 
slopes and zero points at $\log(P) = 1.0$ from the LMC P-L relations, given in Table~\ref{table:compare_slope}.
Since the LMC relations exhibit a smaller dispersion,
we use these slopes to determine the zero point of 
the Galactic relations at $\log(P) = 1.0$.
Following \citet{monson12}, the apparent distance moduli are determined by differencing the LMC and the Galactic zero points.
These distance moduli, presented in Table~\ref{table:results_mu}, are found to be consistent with the recent 
studies on distance determination \citep{fouque07, monson12, piet13}. 
All these results provide an average value of the LMC distance modulus $\mu_{\rm LMC}=18.47\pm0.07$~mag, which is in 
excellent agreement with the ``concordance'' distance modulus of $\mu_{\rm LMC}=18.49\pm0.09$~mag estimated by \citet{delmc14}.

We note that the LMC distance moduli estimated using the $J$-band P-L and the $W_{J,H}$ relations show the largest 
deviations from the other estimates and the \citet{piet13} value. Since the slope and intercepts are nearly 
equal for both Galaxy and LMC, we investigate the possible reasons for the difference. We find that the Galactic 
$J$-band P-L relation and $W_{J,H}$ Wesenheit show a break around 10 days. We use the F-test \citep{anupam14} to 
determine the significance of these breaks and find that the $W_{J,H}$ Wesenheit
is significantly non-linear. The LMC P-L and P-W relations were previously found to be non-linear at 10 days 
\citep{tammann04, ccn05, choong06, varela13}. A detailed statistical analysis of the non-linearity in our LMC 
relations and its impact on the distance scale will be presented in a subsequent study.

\section{A Distance to the Andromeda Galaxy (M31)}
\label{sec:m31plr}

\begin{figure*}
\begin{center}
\includegraphics[width=0.75\textwidth,keepaspectratio]{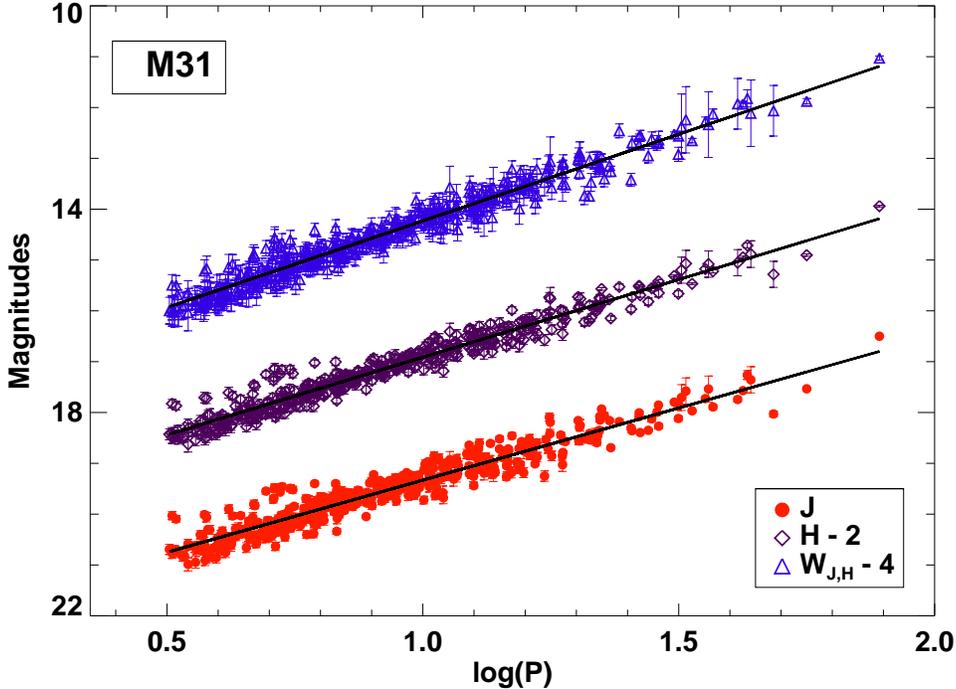}
\caption{NIR P-L and P-W relations for the M31 Cepheids. The solid line represents 
the best fit linear regression to the data points.}
\label{fig:pl_m31.eps}
\end{center}
\end{figure*}

We make use of Cepheid observations in M31 from the Panchromatic Hubble
Andromeda Treasury (PHAT) survey \citep{dal12, william14} to estimate the distance to this galaxy. The observations were carried out 
using the Hubble Space Telescope Advanced Camera for Surveys (HST/ACS) and Wide Field Camera 3 (WFC3).
There are 477 fundamental-mode Cepheids observed with the HST filters F110W and F160W in M31.
Since full light curves are not available, random-phase observations must be used. 
However, the high resolving power of HST allows random-phase observations to be comparable or better 
than ground-based observations. The improved photometric accuracy reduces the dispersion in P-L 
relations even with random-phase magnitudes.

\subsection{NIR P-L and P-W relations}

We note that no robust observational transformation from HST F110W and F160W filters to ground-based $J$ and $H$ 
is available in the literature. Therefore, we make use of theoretical transformations derived from isochrones 
\citep{girardi02}\footnote{\url{http://stev.oapd.inaf.it/cgi-bin/cmd\_2.5}}. 
We take Girardi isochrones over a range of ages (1-12 Gyr) and metallicities (Z=0.0001 to 0.03) at 
$A_{V}=0$ and $A_{V}=1$ \citep{bonatto04, girardi08}. We compare the 2MASS $J$ and $H$ filters to the HST WFC3-IR F110W and 
F160W filters and derive the following transformations over the range of observed F110-F160W colors:

\begin{eqnarray}
J & = & F110W -0.038 - 0.270(F110W-F160W) \nonumber \\
  &   & +~ 0.025(F110W-F160W)^2,\\
H & = & F160W -0.028 - 0.164(F110W-F160W) \nonumber \\
  &   & -~ 0.076(F110W-F160W)^2,
\end{eqnarray}
 
\noindent with {\it rms} errors of $\sim0.012~$mag and $\sim0.011~$mag in $J$ and $H$, respectively. 
We added the {\it rms} error in quadrature 
to the observed photometric error to estimate the associated error for transformed magnitudes. This theoretical transformation
led to an average offset of $0.165$~mag and $0.073$~mag between $J$ \& F110W and $H$ \& F160W, respectively.

The random-phase magnitudes are corrected for reddening using the extinction law 
of \citet{card89} using $R_{V}$ = 3.1 and a foreground reddening to M31 of $A_V=0.17$~mag \citep{schlafly11}.
We derive the P-L relations in $J$ and $H$ and the P-W relation in $W_{J,H}$ using 
the transformed magnitudes. We calculate Wesenheit magnitudes using 
equation~\ref{eq:pw_all} and remove $3\sigma$ outliers and fit the remaining sample of 440 stars to derive
P-L relations and a $W_{J,H}$ P-W relation. These relations are plotted in Fig.~\ref{fig:pl_m31.eps},
while their intercepts and slopes are given in Table~\ref{table:m31_pl}. Our P-L relations in $J$- and
$H$-band are consistent with P-L relations in HST filters derived by \citet{kodric15}, with slight differences
in slopes presumably due to HST filters to 2MASS transformations. A more detailed comparison of long period 
P-L relations in HST filters with \citet{kodric15} results is provided in \citet{rachel15}.

We also compare the slope of M31 P-L and P-W relations with the Galaxy and LMC. The results of the t-statistical
test are given in Table~\ref{table:m31_mw_lmc}. The slope of the M31 
$J$-band P-L relation is statistically different to the Galactic and LMC P-L relations, while the M31
$H$-band P-L relation shows a slope consistent with the Galactic relation (within the large uncertainty in the latter). 
On the other hand, the M31 $W_{J,H}$ slope is not consistent with our results for the Milky Way or the LMC, 
yet it is in agreement with the results from \citet{inno13}. The possible
reason for this discrepancy may be the random-phase observations in M31 and \citet{inno13} as opposed to magnitudes based on
full-phase light curves for our work. Moreover, the derived theoretical transformations may
also contribute to the difference in P-W relations. 

\begin{table}
\begin{center}
\caption{M31 Cepheid NIR P-L and P-W relations}
\label{table:m31_pl}
\scalebox{1.0}{
\begin{tabular}{ccccc}
\hline
\hline
Band & Slope & Intercept & $\sigma$ &  N   \\
\hline
$J$&     -2.839$\pm$0.040      &    19.331$\pm$0.011      &     0.214& 440 \\
$H$&     -3.056$\pm$0.033      &    18.913$\pm$0.009      &     0.173& 440 \\
$W_{J,H}$&     -3.409$\pm$0.035      &    18.231$\pm$0.010      &     0.183& 440 \\
\hline
\end{tabular}}
\end{center}
\end{table}

\begin{table}
\begin{center}
\caption{Comparison of slopes of the M31 P-L and P-W relations with Galaxy and LMC.}
\label{table:m31_mw_lmc}
\scalebox{1.0}{
\begin{tabular}{cccccc}
\hline
\hline
		&	galaxy	&slope		&Src	&$|T|$	&$p(t)$		\\	
\hline
	$J$&	M31&	-2.839$\pm$0.040	&TW 	&--- &         ---\\
	&	MW&	-3.127$\pm$0.076	&TW 	&      3.44 &      0.00\\      
	&	LMC&	-3.140$\pm$0.004      &M15 	&      9.77 &      0.00\\	
        $H$&	M31&	-3.056$\pm$0.033      &TW 	&--- &         ---\\
	&	MW&	-3.164$\pm$0.074      &TW 	&      1.53 &      0.13\\
	&	LMC&	-3.169$\pm$0.004      &M15 	&      4.39 &      0.00\\
  $W_{J,H}$&	M31&	-3.409$\pm$0.035      &TW 	&--- &         ---\\	
	&    	MW&	-3.223$\pm$0.076      &TW 	&      2.52 &      0.01\\
	&	LMC&	-3.154$\pm$0.014      &TW 	&      7.85 &      0.00\\
	&	LMC&	-3.373$\pm$0.008      &I13 	&      1.53 &      0.13\\	
\hline
\end{tabular}}
{\tablecomments{Source: TW - This work; M15 - \citet{macri15}; I13 - \citet{inno13}.}}
\end{center}
\end{table}

\subsection{The Distance to M31}
We use the $W_{J,H}$~magnitudes for the M31 Cepheids to determine the distance to this galaxy.
Since we have calibrated P-W relations for Galactic Cepheids, 
we can calibrate the absolute Wesenheit magnitudes in $W_{J,H}$ for individual M31 Cepheids.
We use these calibrated absolute magnitudes together with the Wesenheit magnitudes for M31 to find the distance modulus for 
each M31 Cepheid independently. We remove the 
$3\sigma$ outliers in the calculated distance moduli and take the mean value to be the distance modulus to M31. 
However, we note that the P-W relation in $W_{J,H}$ for the Cepheids in M31 is steeper as compared to the Galaxy 
and the LMC. Therefore, we observe a trend as a function of period in the distance moduli for Cepheids in M31.
The mean distance modulus to M31 Cepheids using the Galactic calibration is found to be $\mu_{\rm M31}=24.42\pm0.20$~mag.
Similarly, we also make use of the calibrated P-W relation in $W_{J,H}$ for the LMC Cepheids to determine distance moduli of Cepheids
in M31. We consider an error-weighted mean to find a true distance modulus to M31 
of $\mu_{\rm M31}=24.50\pm0.19$~mag, using the LMC calibration. 

\begin{table}
\begin{center}
\caption{M31 distance moduli}
\label{table:mu_m31_dis}
\scalebox{1.0}{
\begin{tabular}{cccc}
\hline
\hline
Calibrator	&$\mu_{\rm M31}$		&Published	& Source\\
\hline
\multirow{2}{1.5cm}{\centering Galaxy}	&\multirow{2}{2.0cm}{24.42$\pm$0.20}	&24.44$\pm$0.12		&R05\\
						&					&24.36$\pm$0.08		&V10\\
\multirow{2}{1.5cm}{\centering LMC}		&\multirow{2}{2.0cm}{24.50$\pm$0.19}	&24.38$\pm$0.06		&R12\\
						&					&24.46$\pm$0.10		&D14\\
\hline
\multicolumn{4}{c}{Average value = 24.46$\pm$0.20}\\
\hline
\end{tabular}}
{\tablecomments{The values of distance modulus for M31 compiled from literature are taken 
from the sources :  R05 - \citet{ribas05}, V10 - \citet{vila10a}, R12 - \citet{riess12}, D14 - \citet{dem3114}.}}
\end{center}
\end{table}

\begin{table*}
\begin{center}
\caption{The distance moduli to Local Group galaxies using a global fit.}
\label{table:dis_extra}
\scalebox{1.0}{
\begin{tabular}{cccccccccc}
\hline
\hline

	&$N$	&Met.	& 	 \multicolumn{3}{c}{Calibrator}&	\multicolumn{4}{c}{Published}	\\
	&---	&	---	& Galaxy &LMC 		&	Galaxy+LMC&	TRGB		&Ref.	&Cepheid	&Ref.	\\
\hline
     WLM&    29&    7.74&    24.85$\pm$0.11&    24.88$\pm$0.08&    24.92$\pm$0.07&    25.12$\pm$0.15&G11&    24.92$\pm$0.04&G08\\
  IC 1613&    23&    7.86&    24.20$\pm$0.10&    24.22$\pm$0.07&    24.26$\pm$0.07&    24.24$\pm$0.10&G11&    24.29$\pm$0.04&P06\\
      SMC&    602&    7.98&    18.96$\pm$0.08&    19.00$\pm$0.05&    19.03$\pm$0.05&    18.98&R07&    18.96$\pm$0.02&D15\\
   NGC 55&    36&    8.05&    26.34$\pm$0.09&    26.35$\pm$0.06&    26.37$\pm$0.06&    ---&---&    26.43$\pm$0.04&GI8\\
 NGC 3109&    69&    8.06&    25.45$\pm$0.09&    25.47$\pm$0.06&    25.49$\pm$0.06&    25.42$\pm$0.13&G11&    25.57$\pm$0.02&S06\\
 NGC 6822&    20&    8.14&    23.39$\pm$0.08&    23.41$\pm$0.06&    23.43$\pm$0.06&    23.26$\pm$0.10&G11&    24.38$\pm$0.02&R14\\
  NGC 300&    15&    8.35&    26.26$\pm$0.10&    26.28$\pm$0.07&    26.29$\pm$0.07&    26.48$\pm$0.04&R07&    26.37$\pm$0.05&G05\\
  NGC 247&    10&    ---&    27.57$\pm$0.12&    27.58$\pm$0.09&    27.60$\pm$0.09&    ---&---&    27.64$\pm$0.04&G09\\
      M33&    24&    8.55&    24.60$\pm$0.08&    24.61$\pm$0.06&    24.62$\pm$0.06&    24.71$\pm$0.04&R07&    24.62$\pm$0.07&G13\\
\hline
       $b_w$&&&            -5.980$\pm$0.072&    -6.009$\pm$0.050&    -6.010$\pm$0.049&&&&\\
   $M_{w,1}$&&&            -3.238$\pm$0.027&    -3.249$\pm$0.019&    -3.244$\pm$0.016&&&&\\
\hline
\end{tabular}}
{\tablecomments{The metallicity ($12+\log[O/H]$) values are taken from \citet{sakai04, bono10, fioren12}. The published values of 
distance moduli are taken from the sources : G05 - \citet{ngc300}, G06 - \citet{ngc6822}, S06 - \citet{ngc3109}, P06 - \citet{ic1613}, 
R07 - \citet{rizzi07}, G08 - \citet{wlm}, GI8 - \citet{ngc55}, G09 - \citet{ngc247}, G11, \citet{trgb11}, F12 - \citet{fiest12}, 
G13 - \citet{m33}, R14 - \citet{rich14}, D15 - \citet{deg15}.}}
\end{center}
\end{table*}

These results are again consistent with previous studies \citep{stan98, ribas05, vila06, vila10a, riess12, vallas13}. 
The values of mean distance modulus for M31 Cepheids obtained using both Galaxy and LMC as calibrators are given in 
Table~\ref{table:mu_m31_dis}. The larger error in distance moduli for M31 can be attributed to a greater scatter in the random-phase 
P-L relation obtained from the single epoch observations from the PHAT survey. However, our results are still in excellent 
agreement with the ``concordance'' distance modulus of $\mu_{\rm M31}=24.46\pm0.10~$mag from \citet{dem3114}. We also note that
\citet{rachel15} determined a distance of $24.51\pm0.08$~mag to M31 using long-period ($P>10~$d) Cepheids and the P-W relation in
HST filters.

\section{Distances to Local Group Galaxies}

We compiled published NIR mean magnitudes for Cepheids in other Local Group galaxies.  
Recently, \citet{ngeow15} derived the P-L relations for Cepheids in SMC at multiple wavelengths. They
used the 2MASS counterparts of OGLE-III SMC fundamental-mode Cepheids and applied random phase corrections to obtain mean $JHK_s$
magnitudes. Also, \citet{rich14} determined the distance to NGC$\,$6822 using previously-published and newly-obtained 
data in multiple bands. The $JHK_{s}$ band photometry was calibrated to the 2MASS system. We make use of NIR 
$J$ and $K_s$ mean magnitudes from these studies in our analysis. The Cepheids in IC$\,$1613 were observed by 
\citet{vicky1613} using the FourStar NIR camera at Las Campanas and the mean magnitudes are available in $JHK_{s}$ bands. 
We also use $J$ \& $K$ observations from the Araucaria project for Cepheids in IC$\,$1613, M33, WLM, NGC$\,$3109,
NGC$\,$300, NGC$\,$55, NGC$\,$247 \citep{ic1613, m33, wlm, ngc3109, ngc300, ngc55, ngc247}. All these mean magnitudes are transformed to 
the 2MASS system using color transformations as discussed in previous sections.

We determine the distance moduli to these Local Group galaxies using the $W_{J,K_s}$ P-W relation. 
We prefer the P-W relations as they are independent of extinction corrections. We use a global fit
to all Cepheids in the Local Group galaxies having $W_{J,K_s}$ Wesenheit magnitudes. Therefore, the 
Wesenheit magnitude $W_{i,j}$ for the $j^{th}$ Cepheid in $i^{th}$ target galaxy is defined as:

\begin{equation}
W_{i,j} = \mu_{i} + M_{w,1} + b_{w}\log P_{i,j}
\end{equation}

\noindent where $\mu_{i}$ is the distance moduli to 
the target galaxy, and $M_{w,1}$ is the Wesenheit magnitude of a Cepheid with $P=10$d in
the calibrator galaxy (LMC and/or Milky Way). The parameter $b_{w}$ is to be determined using the 
global fit and represents the slope for all Cepheids in the sample. We solve the matrix
equation $y=Lq$ using the minimization of $\chi^2$ as discussed in \citet{riess09}. We use $W_{J,K_s}$ 
magnitudes for the Galaxy and LMC separately in the above equation to determine distances to other galaxies.
We also use a combined calibration based on Galactic and LMC data.
The metallicity gradients of Local Group galaxies are based on the $T_{e}$ scale and adopted 
from \citet{sakai04, bono10, fioren12}. We apply the metallicity corrections for calibrations based on the
Galaxy and LMC such that $\mu_{i,0} = \mu_{i} + \gamma(\Delta\log[O/H])$, where
$\Delta\log[O/H]$ is the difference in mean metallicity between the target and the calibrator galaxy 
and $\gamma=-0.05\pm0.06$~mag~dex$^{-1}$ is adopted from \citet{bono10} for $W_{J,K_s}$.
The mean metallicity values in this scale for the Galactic and LMC Cepheids are $8.60$ and $8.34$~dex, respectively. 
However, we do not apply a metallicity correction when we use the combined Galactic+LMC calibration 
in the global fit. The estimated values of the distance moduli are presented in Table~\ref{table:dis_extra}.
The uncertainties in the distance moduli obtained from the global fit are only statistical; 
we also add in quadrature the systematic uncertainty in the zeropoint of the calibrator relations to arrive at the final values.

\begin{figure}
\begin{center}
\includegraphics[width=1.0\columnwidth,keepaspectratio]{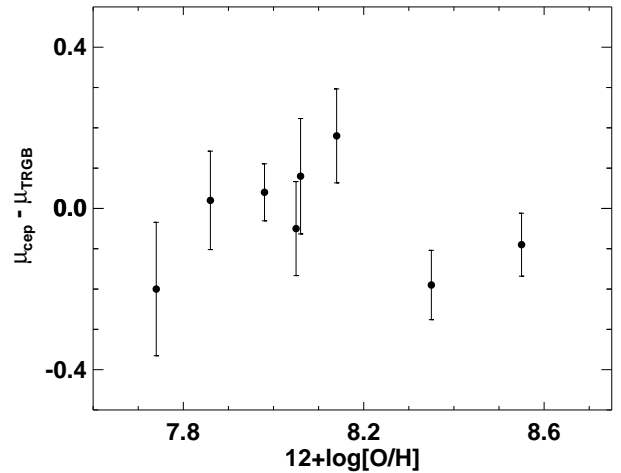}
\caption{The comparison of Cepheid and TRGB distances to Local Group galaxies as a function of metallicity.}
\label{fig:mu_met.eps}
\end{center}
\end{figure}

We note that the distance moduli obtained for IC$\,$1613 are in good agreement with the results based on P-L relations 
by \citet{ic1613} and \citet{vicky1613}. However, there is a large offset ($\sim 0.2$~mag) in the $K_s$ 
magnitudes for Cepheids in common between these two studies. Using the \citet{ic1613} data for our 
P-W analysis yields a distance modulus consistent with previous work, indicating a problem with 
the calibration of the \citet{vicky1613} data. We compare our results with recent TRGB and Cepheid distances 
available in the literature and find a good agreement. The difference in Cepheid and TRGB
based distance estimates as a function of metallicity is shown in Fig.~\ref{fig:mu_met.eps}.
We do not observe any significant trend in estimated distances as a function of metallicity.
Further, the metallicity correction leads to a
difference of $\sim0.06$~mag in distance modulus for metal poor galaxies (WLM, IC$\,$1613, SMC),
while the mean difference is $\sim0.03$~mag with or without metallicity correction for other Local 
Group galaxies. The global fit results in a universal slope of 
-$3.244\pm0.016$ for the $W_{J,K_s}$ Wesenheit relation for Cepheids in Local Group galaxies.
We also note that our distance estimates are consistent for 
a large metallicity range $7.7<12+\log[O/H]<8.6$~dex and therefore, our calibrator relations can be
applied to future observations of Cepheids in more distant galaxies.

\section{Conclusions}
\label{sec:discuss}

In the present analysis, we analysed Period-Luminosity and Period-Wesenheit relations for 
Cepheids in the LMC, the Galaxy, and M31 at $JHK_s$ wavelengths. We also determine the distances to LMC, M31 and other Local
Group galaxies. We summarize our conclusions arising from this study -

\begin{enumerate}

\item{We use $JHK_s$ data for Cepheids from LMCNISS \citep{macri15} to derive new 
Period-Wesenheit relations at these wavelengths. The relations for fundamental-mode Cepheids are based on a sample
size 9 times larger than the previously-published time-series results. The first-overtone Period-Wesenheit relation is calibrated for
the first time with phased light curve data, as opposed to random single-phase observations.}

\item{We obtain a new calibration of Galactic Cepheid Period-Luminosity and Period-Wesenheit relations based on distances
from various methods, taking into account the intrinsic scatter of each technique. Our results bridge the inconsistency between 
Galactic P-L relations based on independent distances and P-L relations derived using Wesenheit distances. We find our 
results are consistent with most of the previously-published work, considering the large intrinsic scatter in Galactic relations.}

\item{We use the new LMCNISS data to provide an independent estimate of the distance to the LMC. Using Galactic calibrations,
we determine $\mu_{\rm LMC} = 18.47$, with a total statistical uncertainty of $\pm0.07$~mag, which is in excellent 
agreement with the value from \citet{piet13} based on late-type eclipsing binaries. However, our error estimates
do not include the unknown systematic uncertainties.}

\item{We derive new P-L and P-W relations for Cepheids in M31, based on the observations from the PHAT survey. We
develop theoretical transformations from HST filters F110W and F160W to 2MASS $J$ and $H$-bands. Although the relations
are based on random-phase observations, the highly accurate HST observations help to reduce the observed dispersion in
Period-Luminosity and Period-Wesenheit relations.}

\item{Using Galactic and LMC $W_{J,H}$ Wesenheit relations as reference, we estimate a distance modulus for M31 of
$\mu_{\rm M31} = 24.46\pm0.20$~mag, in excellent agreement with recent determinations \citep{riess12, vallas13, dem3114}.} 

\item{We apply a simultaneous fit to Cepheids in Local Group galaxies, using the Galaxy and LMC as calibrators, to
obtain a global slope of -$3.244\pm0.016$~mag/dex in $W_{J,K_s}$ and estimate robust distances, which are found
to be consistent with previous results based on TRGB and Cepheids. We do not find a significant metallicity effect at these wavelengths.}

\item{Our absolute calibration of the Galactic and LMC relations provides accurate distances for Local Group galaxies with a wide metallicity
range ($7.7<12+\log[O/H]<8.6$)~dex. In combination with higher-quality NIR light curves for Cepheids at greater distances,
they can be used for further improvements in the precision and accuracy of the distance scale.}
\end{enumerate}

An upcoming study based on LMCNISS data (Bhardwaj et al. in preparation) will include a statistical analysis of non-linearities
in the Leavitt law at $VIJHK_s$ wavelengths and its impact on the distance scale 
and in constraining theoretical pulsation models. 

\section*{Acknowledgments}
\label{sec:ackno}

AB acknowledges the Senior Research Fellowship grant 09/045(1296)/2013-EMR-I
from Human Resource Development Group (HRDG), which
is a division of Council of Scientific and Industrial Research
(CSIR), India. This work is supported by the grant for 
the Joint Center for Analysis of Variable Star Data
provided by Indo-U.S. Science and Technology Forum.
LMM acknowledges support by the United States 
National Science Foundation through AST grant number 1211603 and by Texas A\&M 
University through a faculty start-up fund and the Mitchell-Heep-Munnerlyn 
Endowed Career Enhancement Professorship in Physics or Astronomy.
CCN thanks the funding from Ministry of Science and Technology (Taiwan) under
the contract NSC101-2112-M-008-017-MY3. 
This work also makes use of data products from the 2MASS survey, 
which is a joint project of the University of Massachusetts and the Infrared Processing
and Analysis Center/California Institute of Technology, funded by the National Aeronautics 
and Space Administration and the National Science Foundation.
In addition, this study also makes use of NASA's Astrophysics Data System, 
the VizieR catalogue and the McMaster Cepheid Photometry data base.

\appendix

\section{A Comparison of Period-Luminosity and Period-Wesenheit relations}
\label{sec:comp_mw_lmc}

We use our LMCNISS $JHK_s$ mean magnitudes to determine a distance to the LMC with published NIR Galactic P-L
relations listed in Table~\ref{table:compare_slope}. We present the LMC distance moduli obtained using 
these P-L relations in Table~\ref{table:mu_mw}. We note that the distances in $H$ and $K_s$ bands are
considerably smaller using P-L relations from \citet{fouque07} and \citet{storm11}. Similarly, the $J$ band P-L from
\citet{ngeow12} leads to a relatively greater value of LMC distance as compared to \citet{piet13}. We 
explore the reasons of possible discrepancy among these relations and compare with our Galactic P-L
relations derived in the present study.

\citet{ngeow12} used a method involving the Wesenheit function to derive distance moduli 
for a large number of Galactic Cepheids and found a marginal average difference (-0.061 to 0.009) with published distances. 
This method was also calibrated against HST parallaxes but the uncertainties listed in that work are only statistical errors.
It is important to note that even though the Wesenheit distances are consistent with other methods, there is a significant 
change in the slope and intercepts of P-L relations from \citet{ngeow12} with \citet{fouque07} and \citet{storm11}. 
Interestingly, our results based on various 
distances are very consistent with \citet{ngeow12}.

We find that our slopes for $JHK_s$ P-L relations are consistent with 
\citet{fouque07}, \citet{storm11} and \citet{ngeow12} as $p(t)>0.05$, in
all the bands. However, the intercepts of P-L relations show mixed results with most of them being consistent with published work.
The intercepts of $JHK_s$ P-L relations are in excellent agreement with \citet{ngeow12} but relatively smaller 
than \citet{fouque07} and \citet{storm11}. The t-test suggests that the zero-points of our P-L relations are statistically different
to \citet{fouque07} but the zero-point of the $H$ and $K_s$-band P-L relations are statistically similar to 
\citet{storm11}, with $J$-band zero-point again being significantly different. We also note that the dispersion 
in our P-L relations is similar to those by \citet{storm11} whereas we have increased the sample size nearly 1.5 times.
The discrepancy in results with 
\citet{fouque07} is mainly due to significantly different sample sizes.

We test the difference in zero-points with 
\citet{fouque07} and \citet{storm11} by comparing the properties of P-L relations derived using only Cepheids common to these samples.
We find that the difference in zero-points of the
two set of P-L relations is reduced on an average by $0.02$~mag. Therefore, the slope and intercepts of our P-L relations
are not significantly different from published work.
A small contribution to this difference in intercepts may be due to the inclusion of few first overtone 
stars (for example, FN AQL, V496 AQL and Y OPH) in \citet{fouque07} and \citet{storm11}. These stars are 
not considered in our sample following 
\citet{ngeow12}. Our results for the P-W relation in $W_{J,K_s}$ are also consistent with the findings of \citet{storm11}.

\begin{table}
\begin{center}
\caption{Comparison of LMC distances using the published Galactic P-L relations.}
\label{table:mu_mw}
\scalebox{1.0}{
\begin{tabular}{cccc}
\hline
\hline
Source &$J$	&$H$	&$K_s$\\
\hline
F07&      18.44$\pm$      0.05&      18.32$\pm$      0.05&      18.37$\pm$      0.06\\
S11&      18.40$\pm$      0.06&      18.38$\pm$      0.06&      18.40$\pm$      0.07\\
N12&      18.56$\pm$      0.05&      18.47$\pm$      0.05&      18.50$\pm$      0.05\\

\hline
\end{tabular}}
{\tablecomments{ The source column represents the calibrator P-L relations from : F07 - \citet{fouque07}, S11 - \citet{storm11}, 
N12 - \citet{ngeow12}.}}
\end{center}
\end{table}

\begin{table}
\begin{center}
\caption{Comparison of calibrated Galactic and LMC P-L and P-W relations derived in the present study.}
\label{table:mw_lmc}
\scalebox{1.0}{
\begin{tabular}{cccccccc}
\hline
\hline

		&	&$J$	&	$H$&	$K_{s}$& 	$W_{J,H}$&	$W_{J,K_{s}}$&	$W_{H,K_{s}}$\\	 
\hline
Slope		&$|T|$	&0.27	&0.12	&	0.18 &		1.59 &		3.03 &		3.35 \\	
		&$p(t)$	&0.79	&0.90	&	0.86 &		0.11 &		0.00 &		0.00 \\
Intercept	&$|T|$	&0.20	&0.49	& 	0.20 &		1.24 &  	0.44 &		0.09 \\
		&$p(t)$	&0.84	&0.63	&	0.84 &		0.21 &		0.66 &		0.92 \\
\hline
\end{tabular}}
\end{center}
\end{table}

We also compare the slopes and intercepts of Milky Way and LMC Cepheid P-L and P-W relations. From Table~\ref{table:lmc_slope}
and \ref{table:compare_slope}, we find that the intercepts of both P-L and P-W relations for the Galaxy and the LMC are 
essentially similar in all the bands. The t-test results, given in Table~\ref{table:mw_lmc}, also provide evidence of
statistically equal zero-points under a $95\%$ confidence level. Further, the slopes of the P-L and P-W relations for 
both the Galaxy and LMC are also very similar except in $W_{J,K_{s}}$ and $W_{H,K_{s}}$. This difference in the
slopes of the Wesenheit relations is mainly due to insignificant contribution of color terms in Galactic Wesenheits,
which leads to greater dispersion than P-L relations. This provides further empirical evidence that at NIR 
wavelengths, P-L and P-W relations for Cepheids are universal and the zero-points are independent of 
metallicity effects \citep{mmsai06, fouque07, monson12}.  

\bibliography{cpapir_II}{}
\bibliographystyle{apj}

\end{document}